# Rapid Whole-Heart CMR with Single Volume Super-resolution

Jennifer A. Steeden (PhD)[1],

Michael Quail (MBChB, PhD)[1,2],

Alexander Gotschy (MD)[2,3],

Andreas Hauptmann (PhD) [4,5],

Simon Arridge (PhD)[4],

Rodney Jones (MSc)[1],

Vivek Muthurangu (MD)[1]

1. UCL Centre for Cardiovascular Imaging, University College London, London. WC1N 1EH. United Kingdom
2. Great Ormond Street Hospital, London, WC1N 3JH. United Kingdom
3. Institute for Biomedical Engineering, University and ETH Zurich, Zurich. Switzerland
4. Department of Computer Science, University College London, London. WC1E 6BT. United Kingdom
5. Research Unit of Mathematical Sciences, University of Oulu, Oulu. Finland

**Corresponding Author:** Jennifer Steeden

UCL Centre for Cardiovascular Imaging, Institute of Cardiovascular Science

30 Guildford Street, London. WC1N 1EH

jennifer.steeden@ucl.ac.uk

Tel: +44 (0)207 762 6834

Fax: +44 (0)207 813 8263

**Abstract**

Background: Three-dimensional, whole heart, balanced steady state free precession (WH-bSSFP) sequences provide delineation of intra-cardiac and vascular anatomy. However, they have long acquisition times. Here, we propose significant speed-ups using a deep-learning single volume super-resolution reconstruction, to recover high-resolution features from rapidly acquired low-resolution WH-bSSFP images.

Methods: A 3D residual U-Net was trained using synthetic data, created from a library of high-resolution WH-bSSFP images by simulating 50% slice resolution and 50% phase resolution. The trained network was validated with synthetic test data, comparing the resultant super-resolved data to high-resolution data using mean square error (MSE) and Structural Similarity Index (SSIM). Additionally, prospective low-resolution data and high-resolution data were acquired in 40 patients. Vessel diameters, quantitative and qualitative image quality, were measured on both the low-resolution and super-resolution WH-bSSFP data and compared to high-resolution WH-bSSFP data.

Results: Synthetic low-resolution data had a SSIM of 0.87, and a MSE of $1.28 \times 10^{-3}$, compared to the high-resolution data. After super-resolution reconstruction, the SSIM significantly increased (p<0.05) to 0.96 and the MSE significantly decreased (p<0.05) to $0.68 \times 10^{-3}$. Prospectively acquired low-resolution data was acquired ~x3 faster than the prospective high-resolution data (173s vs 488s). Super-resolution reconstruction of the low-resolution data took <1s per volumes. Qualitative image scores showed super-resolved images had better edge sharpness, fewer residual artefacts and less image distortion than low-resolution images, with similar scores to high-resolution data. Quantitative image scores showed super-resolved images had significantly better edge sharpness than low-resolution or high-resolution images, with significantly better signal-to-noise ratio than high-resolution data. Vessel diameters measurements showed over-estimation in the low-resolution measurements, compared to the high-resolution data. No significant differences and no bias was found in the super-resolution measurements.

Conclusion:

This paper demonstrates the potential of using a residual U-Net for super-resolution reconstruction of rapidly acquired low-resolution whole heart bSSFP data within a clinical setting. We were able to train the network using synthetic training data from retrospective high-resolution whole heart data. The resulting network can be applied very quickly, making these techniques particularly appealing within busy clinical workflow. Thus, we believe that this technique may help speed up whole heart CMR in clinical practice.

## BACKGROUND

Three-dimensional whole heart, balanced steady state free precession (WH-bSSFP) imaging is an important part of the cardiovascular magnetic resonance (CMR) imaging protocol in congenital heart disease (1). This is because WH-bSSFP provides excellent delineation of both intra-cardiac and vascular anatomy. However, WH-bSSFP sequences are usually cardiac triggered and respiratory navigated, resulting in long acquisition times (up to 10 minutes).

Significant speed-ups can be achieved through the use of non-Cartesian sampling (i.e. spiral (2) or radial (3)) or data under-sampling with state-of-the-art reconstruction strategies (i.e. compressed sensing (4)). Unfortunately, these methods require major sequence modifications and are often handicapped by long reconstruction times, even with the use of modern computing (i.e. graphics processing units (5)). An alternative approach is single volume super-resolution reconstruction (SRR), where high-resolution features are recovered from rapidly acquired low-resolution data. The benefits of SRR is that it can be performed as a simple post-processing step without any sequence modification. However, conventional algorithms often produce unrealistic looking images, limiting the utility of this method (6). Recently, machine learning has transformed SRR with the ability to produce realistic high-resolution images from low-resolution data (7-9).

In this study, we use a deep-learning SSR approach to reconstruct high-resolution data from rapidly acquired low-resolution WH-bSSFP images. This was achieved by first creating a 'synthetic' low-resolution training data set from a library of reference standard high-resolution WH-bSSFP images. The paired data were then used to train a convolutional neural network (CNN) to map between low-resolution and high-resolution images (super-resolution). The aims of this study were to: i) Assess the accuracy of deep learning single volume SRR for recovering high-resolution data from synthetically down-sampled WH-bSSFP data, ii) Assess the robustness of the resultant network, at recovering high-resolution data from different resolution input data, iii) Assess the feasibility of using deep learning single volume SRR for reconstruction of prospectively acquired low-resolution WH-bSSFP data, and iv) Compare acquisition time, image quality and accuracy of vessel diameter measurements from single volume SRR, compared to low-resolution and reference standard high-resolution WH-bSSFP images.

**METHODS**

**Network Architecture**

The CNN architecture chosen to perform super-resolution reconstruction in this study was based on a residual U-Net. This architecture has been previously shown to be robust in many applications, such as deep artefact suppression of real-time cine MRI data (10) and ventricular segmentation (11-13). A residual U-Net is a multi-scale CNN where images are sequentially down-sampled and then up-sampled with the network learning the difference between the input and desired output (residual) rather than the desired output directly (14). In a residual U-Net, the learnt residual is added to the input data to produce the final output data (15). In this study, a 3D residual U-Net was trained with paired high-resolution 'ground truth' data and corresponding synthetic low-resolution images (Figure 1). This network structure was chosen as it has been shown to be robust for image reconstruction applications. Each convolutional layer had a filter size of 3x3x3 and was equipped with a rectified linear unit as nonlinearity, except the last layer that produced the residual update. We used a smaller network size than the classical U-Net architecture to avoid overfitting. The filters were equally weighted in all domains and hence no directions were favoured in the training process. The output of the network was projected to positive numbers by a rectified linear unit to enforce non-negativity.

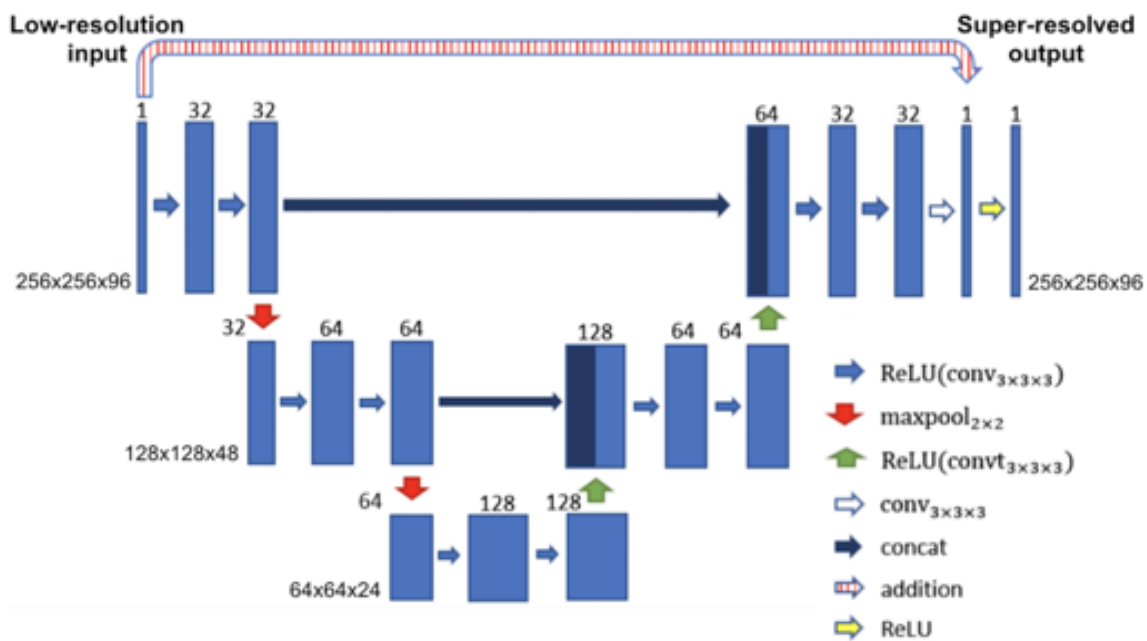

*Figure 1: Network architecture Chosen residual U-Net architecture used for 3D single volume super-resolution. The input is given by the low-resolution WH-bSSFP images. The numbers on top of the blue bars denote the number of channels for each layer. The resolution for each multilevel decomposition is shown on the left. Each convolutional layer is equipped with a Rectified Linear Unit as nonlinearity, given by ReLU(x)=max(x,0).*

**Preparation of Synthetic Training Data**

The synthetic training data was created from conventional high-resolution WH-bSSFP data (without any obvious artefacts due to breathing or arrhythmia) collected from previously scanned children and adults with paediatric heart disease or congenital heart disease. The training data set contained 500 3D WH-bSSFP images (mean age: 26±13 years, range: 5-80 years. Male: 299, Female: 201. Heart rate: 67±9 bpm, range: 41-86 bpm). A full list of diagnoses can be found in Additional File 1. Sequence parameters for the cardiac triggered, respiratory navigated high-resolution WH-bSSFP sequence are shown in Table 1.

|  | High-resolution WH-bSSFP | Low-resolution WH-bSSFP |
|---|---|---|
| Orientation | Sagittal | Sagittal |
| Matrix (*kx-ky*) | ~256x144 | ~256x72 |
| Acceleration in *ky* | x2 (GRAPPA) | x2 (GRAPPA) |
| GRAPPA reference lines | 24 | 24 |
| Partial-Fourier in *ky* | 6/8 | 6/8 |
| FOV *x-y* (mm) | ~400x238 | ~400x238 |
| Slices | ~96 | ~48 |
| Slice thickness (mm) | ~1.6 | ~3.2 |
| Partial-Fourier in *kz* | 6/8 | 6/8 |
| FOV *z* (mm) | ~154 | ~154 |
| Flip angle (deg) | 90° | 90° |
| TE/TR (ms) | ~1.6/~3.6 | ~1.6/~3.6 |
| Bandwidth (Hz/Pixel) | ~592 | ~592 |
| Lines Per segment | ~30 | ~30 |
| Cardiac triggering | Yes | Yes |
| Respiratory navigator | Yes (window 3mm) | Yes (window 3mm) |
| Spatial resolution (mm) | ~1.6x1.6x1.6 | ~1.6x3.2x3.2 |
| Temporal resolution (ms) | ~108 | ~108 |
| Total Acquisition Time (mins) | ~8.1 (range: 3.3-14.8) | ~2.9 (range: 1.1-5.0) |

*Table 1: Imaging parameters*
*Imaging parameters for the training/testing of the network, as well as prospective data.*

Using these 500 data sets, low-resolution data was created by simulating 50% slice resolution and 50% phase resolution. The first step was to crop/pad the high-resolution data to a 256x256 matrix with 96 slices, to make the data consistent for training. This was followed by Fourier transform to produce a synthetic k-space. The outer 50% of k-space in the slice and phase encode direction were then zeroed, simulating two-fold down-sampling of the data in both directions. In addition, 75% partial Fourier in both the slice and phase encoding directions was simulated by further asymmetric zeroing in k-space. The resultant simulated k-space was then inverse Fourier transformed back to image space, and the absolute value taken. This produced the synthetic low-resolution data whilst maintaining a matrix size of 256x256x96. Both the high- and low-resolution whole heart data were further cropped to a 192x192 matrix, in all 96 slices, to constrain the learning problem to the anatomy of interest (heart). Finally, each 3D data set was normalized to have signal intensities in the range [0, 1]. All processing required for creation of the synthetic training data was performed in MATLAB (MATLAB 2016b, The MathWorks, Inc., Natick, Massachusetts, United States). A flow diagram of the steps necessary to create the synthetic data is included in Additional File 2.

**Network Training and Validation**

Implementation and training of the U-Net was done in Python with TensorFlow (16). We minimised the $\ell^1$-loss of the reconstructed volume to the desired ground truth. The training was done for 200 epochs with the Adaptive Moment Estimation algorithm (ADAM) (17), with an initial learning rate of $10^{-3}$ and batches of two volumes. The total training time for each network took ~38 hours on a Titan XP GPU with 12Gb memory.

The trained network was validated with synthetic test data created in the same way as the training data. The test data consisted of 25 previously scanned children and adults with paediatric heart disease or congenital heart disease, not included in the training data set (mean age: 27±12 years, range: 10-51 years. Male: 13. Heart rate: 69±9 bpm, range: 52-85 bpm. A full list of diagnoses can be found in Additional File 1). The resulting super-resolved data were compared to the ground truth, high-resolution data using mean square error (MSE) and Structural Similarity Index (SSIM).

**Generalisability**

The SSR network was specifically trained to super-resolve a given low resolution data set. Therefore, we wanted to assess the robustness of the trained network to inputs with different resolutions of the synthetic down-sampled data. To do this, we used the 25 synthetic test data sets, described above. We simulated resolutions from 10% slice and phase resolution to 100% slice and phase resolution, in increments of 10%. The test data was created as described above, but with varying amount of zeros used in the outer portions of k-space in the slice and phase encode direction. The resulting super-resolved data were compared to the ground truth, high-resolution data using MSE and SSIM. The results of these analyses were averaged over the entire volume for each patient.

**Prospective Clinical Study**

Forty children and adults with paediatric or congenital heart disease referred to our centre for clinical CMR were included in the prospective part of the study during September and October 2019 (mean age: 27±14 years, range: 11-64 years. Male: 20. Heart rate: 68±11 bpm, range: 45-95 bpm. A full list of diagnoses can be found in Additional File 1). Exclusion criteria were: i) Significant metal artefact due to implanted medical devices, and ii) Arrhythmia. All patients were imaged on a 1.5T MR scanner (Avanto, Siemens Healthcare, Erlangen, Germany) with vector electrocardiographic (VCG) gating. Low-resolution WH-bSSFP and high-resolution WH-bSSFP data were both acquired on all subjects (see Table 1 for acquisition parameters). The trained network was then used to perform super-resolution reconstruction on the low-resolution data.

The use of retrospectively collected training and test data, as well as collection of prospective whole heart data was approved by the local research ethics committee, and written consent was obtained from all subjects/guardians (Ref: 06/Q0508/124).

**Analysis of Prospective Data**

Vessel diameters, as well as quantitative and qualitative image quality, were measured on both the low-resolution (LR) and super-resolution (SR) WH-bSSFP data and compared to reference standard high-resolution (HR) WH-bSSFP data. All measurements were made using in-house plugins for the OsiriX open source DICOM viewing platform (Osirix v.9.0, OsiriX foundation, Switzerland) (18).

*Vessel Diameter Measurements*

Diameters were measured by two CMR specialists (M.Q. and A.G.) from multi-planar reformats (MPR's) of the ascending aorta (AAo), descending aorta (DAo), main pulmonary artery (MPA), right pulmonary artery (RPA), and left pulmonary artery (LPA). Each clinician was the primary observer for 20 unique patient data sets, of which 10 were re-evaluated to assess intra-observer variability. In addition, each observer assessed 10 patient data sets from the other primary observer, to evaluate inter-observer variability. Thus, each observer scored and processed 40 patient data sets. Overall 20 patient data sets were used to evaluate intra-observer variability and the other 20 patient data sets used to evaluate inter-observer variability. All images were viewed in a randomised order. For each vessel, two perpendicular diameter measurements were made, and the average was used for all further analyses.

*Qualitative and Quantitative Image Quality*

All MPR data was graded on a 5-point Likert scale in three categories: sharpness of vessel borders (1 = non-diagnostic, 2 = poor, 3 = adequate, 4 = good, 5 = excellent), image distortion (1 = non-diagnostic, 2 = severe, 3 = moderate, 4 = mild, 5 = minimal), and residual artefacts (1 = non-diagnostic, 2 = severe, 3 = moderate, 4 = mild, 5 = minimal).

Vessel edge sharpness (ES) was also calculated from MPR's by measuring the maximum gradient of the normalized pixel intensities across the border of the vessel of interest as previously described (19). Edge sharpness was calculated in sixty positions around the vessel, and the average value was used for comparison.

Estimated signal-to-noise ratio (SNR) and contrast-to-noise ratio (CNR) were assessed in a mid-thoracic slice that included blood pool, ventricular myocardium and lung. SNR was calculated as the ratio of average blood signal intensity to the average noise signal intensity, taken in the lungs (20). CNR was calculated as the ratio of blood signal intensity to average myocardial signal intensity (20).

**Statistics**

Statistical analyses were performed by using the R software (Rstudio, v.3.5). Comparisons of continuous variables (vessel diameters, edge sharpness, SNR and CNR) across of all three groups was performed using one-way repeated measures analysis of variance (ANOVA) with post hoc testing using Holm correction for significant results. Comparison of Likert data was performed using the Friedman's test with post-hoc testing using the Nemenyi test for significant results. Inter and intra-observer variability was assessed using one-way intraclass correlations (ICC), displayed with their 95% confidence intervals. Comparison of acquisition time between the high-resolution and low-resolution WH-bSSFP sequences was performed using a paired t-test. For assessment of agreement of diameter measurements, the high-resolution WH-bSSFP data was used as the reference standard for Bland-Altman analysis. A p-value of less than 0.05 indicated a significant difference.

**RESULTS**

**Network Validation**

Figure 2 shows examples of original high-resolution data, simulated low-resolution data and accompanying super-resolved data. Due to the simulated down-sampling, the low-resolution data had a SSIM of 0.87±0.02, and a MSE of 1.28±0.57 $\times 10^{-3}$, compared to the high-resolution data. After super-resolution reconstruction, the SSIM significantly increased ($p<0.05$) to 0.96±0.01 and the MSE significantly decreased ($p<0.05$) to 0.68±0.45 $\times 10^{-3}$. This demonstrates that super-resolution reconstruction enables recovery of features lost in the low-resolution simulation. Additional File 3 shows the same synthetic tests, as trained with alternate network structures, demonstrating the residual U-net, with an $\ell^1$-loss function gave the best results.

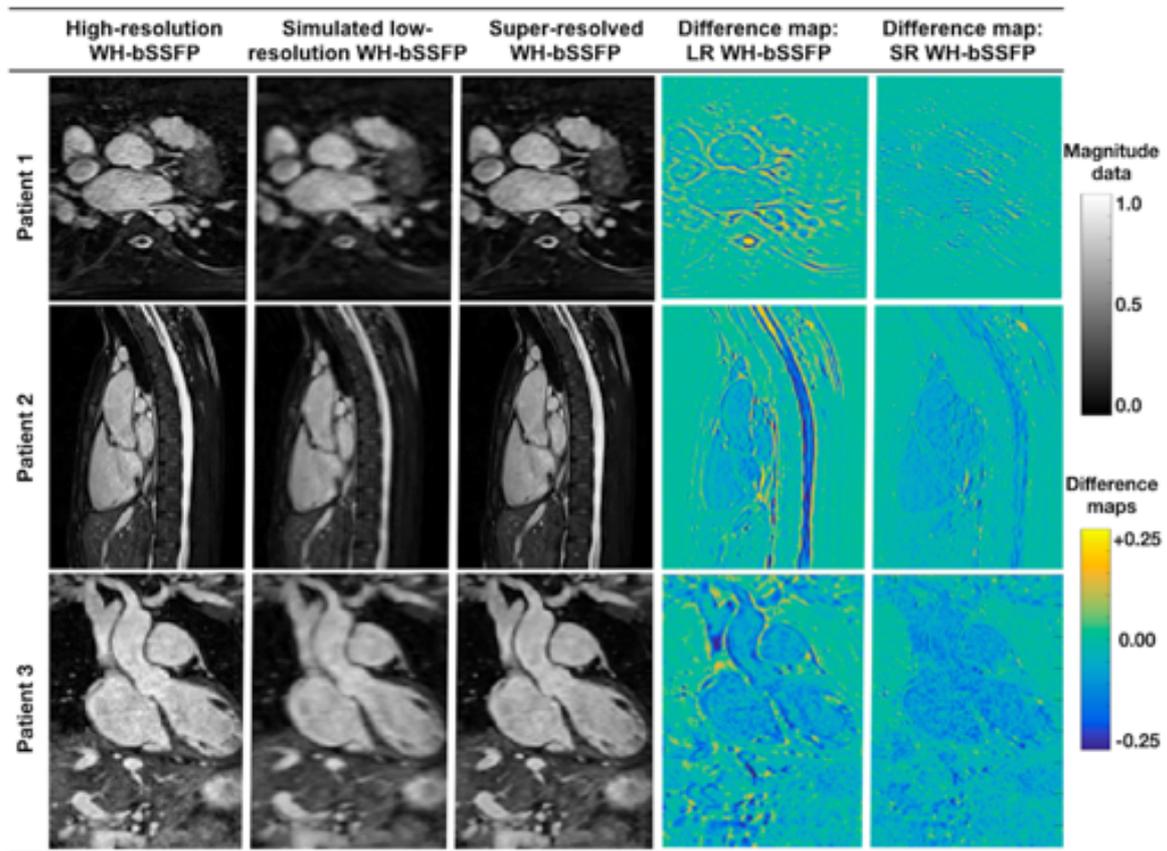

*Figure 2: Synthetic test data*

*Example image quality from the synthetic test data in three patients. From left to right: Original high-resolution WH-bSSFP data; Simulated low-resolution WH-bSSFP data; Resulting super-resolved data; Difference between high-resolution data and simulated low-resolution data; Difference between high-resolution data and super-resolved data.*

**Generalisability**

Figure 3a and 3b show that SSIM is highest and MSE is lowest when the input data has the same resolution as the data used for training (50% phase and slice resolution. This can be seen visually in Figure 3c – at lower resolutions, the network is unable to recover high resolution features resulting in significantly blurred images.

At higher resolutions, the network created artificially sharp edges in the resultant images (Additional File 4 shows a table of the results).

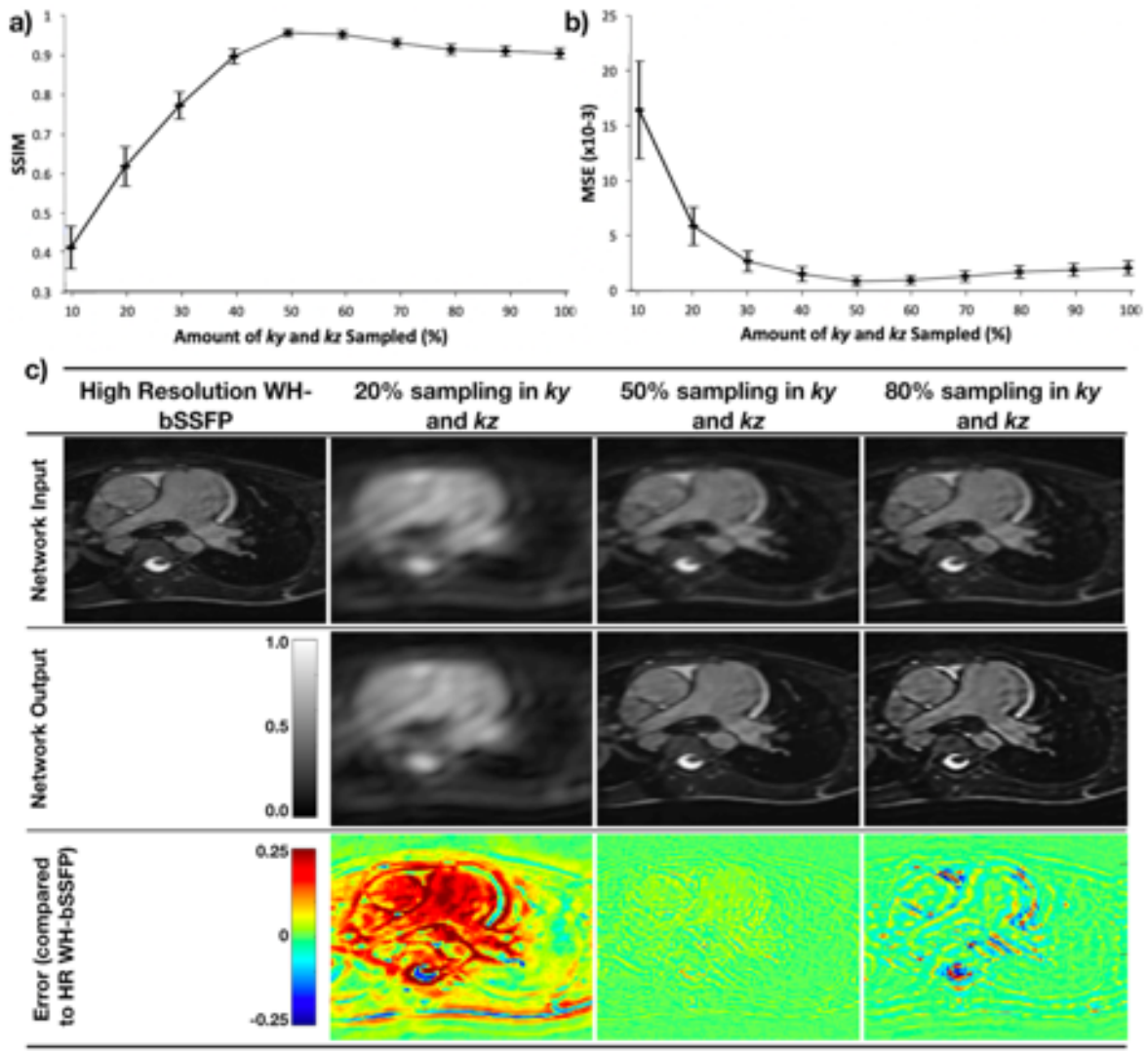

*Figure 3: Generalisability tests* Results from the generalisability tests performed on 25 synthetic test data sets. Agreement of super-resolved images with the reference high-resolution WH-bSSFP images at different amounts of down-sampling of the input data; a) SSIM, b) MSE. c) Example low-resolution images at different amounts of down-sampling (input to network), the super-resolved results from the network, and the error maps comparing the super-resolved images to the truth images. See Additional File 4 for full results.

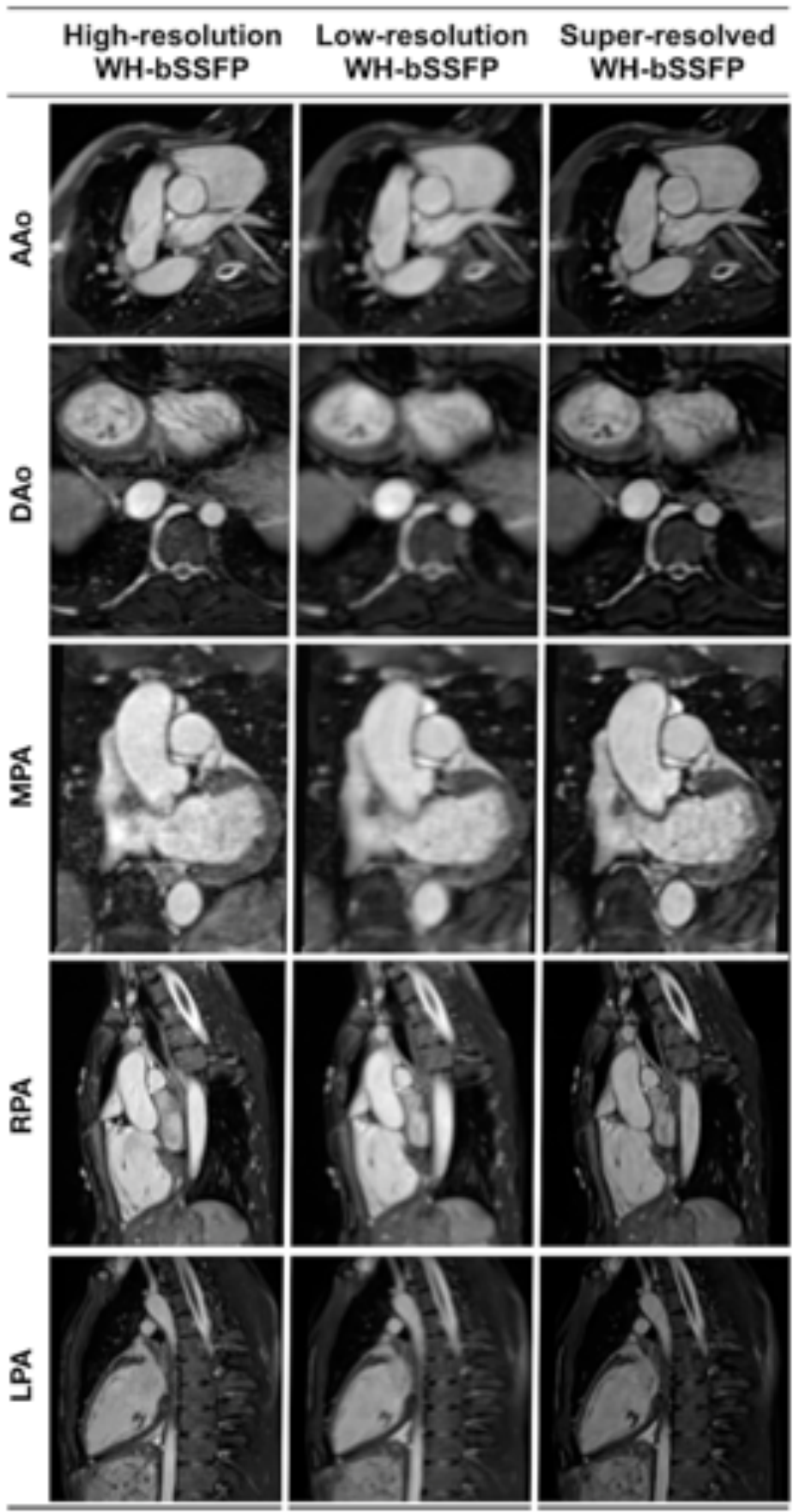

*Figure 4: Example images of vessels from the prospective study*
Representative image quality from the prospective study. Multi-planar reformats of the ascending aorta (AAo), descending aorta (DAo), main pulmonary artery (MPA), right pulmonary artery (RPA), and left pulmonary artery (LPA), from the high-resolution and low-resolution acquisitions, as well as the super-resolved result.

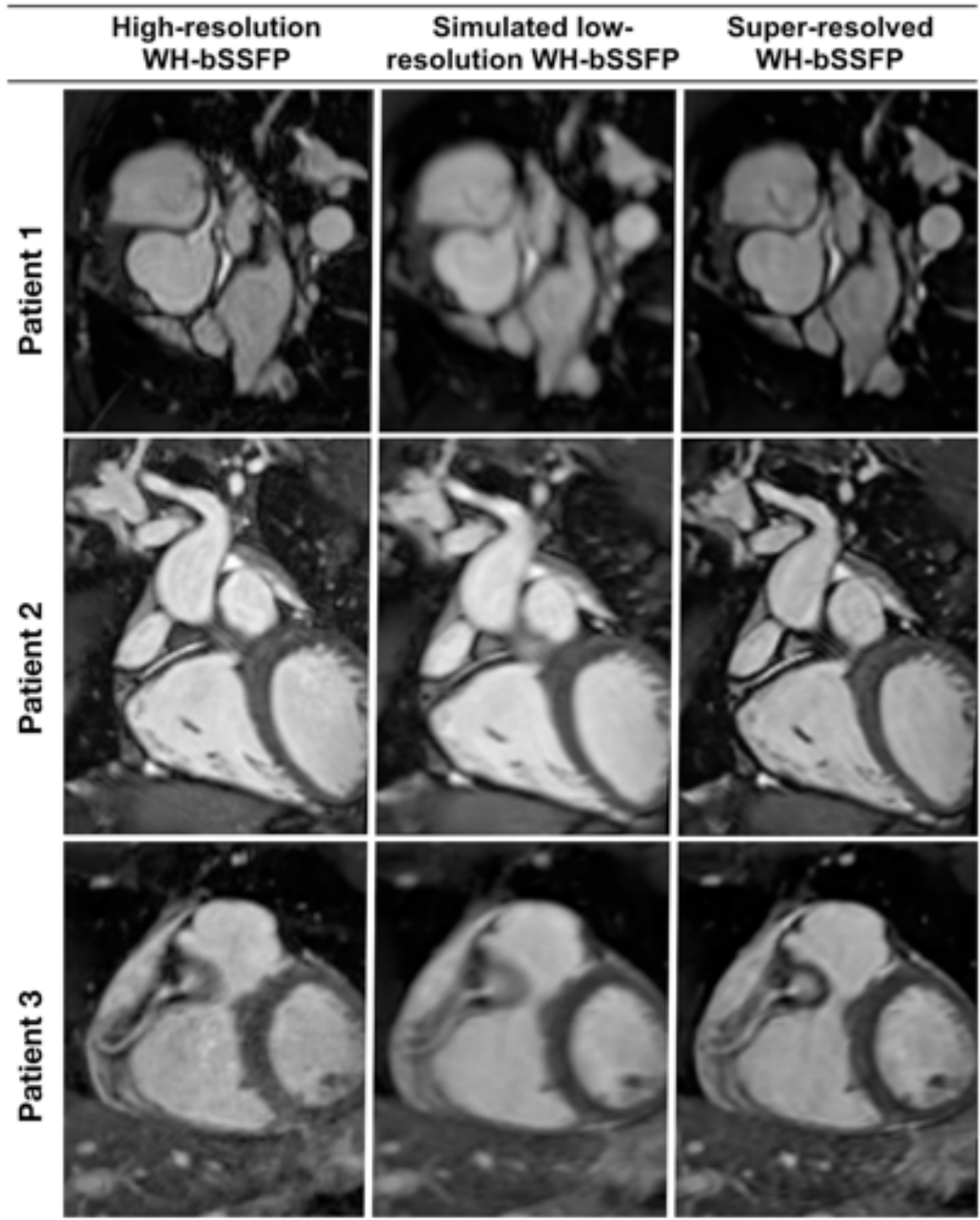

***Figure 5: Example images of the coronaries from prospective study***

*Representative image quality from the prospective study. Multi-planar reformats of the coronary artery from the high-resolution and low-resolution acquisitions, as well as the super-resolved result.*

**In-vivo Study**

High-resolution and low-resolution WH-bSSFP data were successfully acquired in all 40 patients. Total acquisition time for HR-WH data (488±138 s, range: 200 to 889 s) was significantly ($p<0.05$) higher than the LR-WH data (173±54 s, range: 66 to 302 s). The average speed-up in acquisition time was x2.9±0.8 (range: 1.5 to 5.4).

SSR was successfully applied to all low-resolution WH-bSSFP data sets. The network took ~0.7 seconds to perform super-resolution per volume (on a Titan XP GPU with 12Gb memory). Representative images are shown in Figure 4 and Figure 5. It can be seen that image sharpness is improved between the low-resolution data and the super-resolution reconstruction. This is particularly evident in small vessels, such as the coronary arteries (Figure 5).

*Quantitative Vessel Diameter Measurements*

Vessel diameters measured from high-, low- and super-resolution data are shown in Table 2. Figure 6 shows the Bland-Altman plots for all vessels combined, with the Bland-Altman plots for each of the individual vessels shown in Additional File 5. A small but significant overestimation was found in the ascending aorta, descending aorta and right pulmonary artery diameters using the low-resolution data, and a trend for overestimation in the main pulmonary artery diameter. There were no significant differences between the high-resolution and super-resolution data.

| Vessel | n | Mean Diameter ± Standard deviation (mm) | | | Bias (Limits of Agreement) | |
|---|---|---|---|---|---|---|
| | | *High-resolution* | *Low-resolution* | *Super-resolution* | *Low-resolution* | *Super-resolution* |
| AAo | 40 | 28.0 ± 5.6 | 28.5 ± 5.6* | 28.0 ± 5.7$^\dagger$ | 0.5 (-1.7 to 2.6) | -0.05 (-2.2 to 2.1) |
| DAo | 40 | 17.2 ± 2.7 | 17.7 ± 3.0* | 17.2 ± 2.7$^\dagger$ | 0.6 (-0.8 to 1.9) | 0.01 (-1.1 to 1.2) |
| MPA | 40 | 24.1 ± 3.7 | 24.6 ± 3.7 | 24.4 ± 3.7 | 0.4 (-1.8 to 2.7) | 0.2 (-2.1 to 2.6) |
| RPA | 40 | 16.2 ± 3.3 | 16.6 ± 3.4* | 16.3 ± 3.4$^\dagger$ | 0.4 (-1.2 to 2.0) | 0.08 (-1.9 to 2.1) |
| LPA | 40 | 17.3 ± 3.2 | 17.3 ± 3.2 | 17.0 ± 2.9 | 0.08 (-2.0 to 2.1) | -0.2 (-2.3 to 1.8) |
| Overall | 200 | 20.6 ± 6.0 | 20.9 ± 6.1 | 20.6 ± 6.1 | 0.4 (-1.5 to 2.3) | 0.008 (-2.0 to 2.0) |

*\* Indicates significant differences with standard high-resolution WH-bSSFP technique as assessed by ANOVA with post hoc testing using Holm correction (p<0.05)*

*$^\dagger$ Indicates significant differences with low-resolution WH-bSSFP technique as assessed by ANOVA with post hoc testing using Holm correction (p<0.05)*

*Bias is the mean of the paired difference with the high-resolution WH-bSSFP*

*Limits of agreements are bias ± 1.96xSD*

*Table 2: **Vessel diameter measurements** Vessel diameter measurements from the prospective patient study (primary observer).*

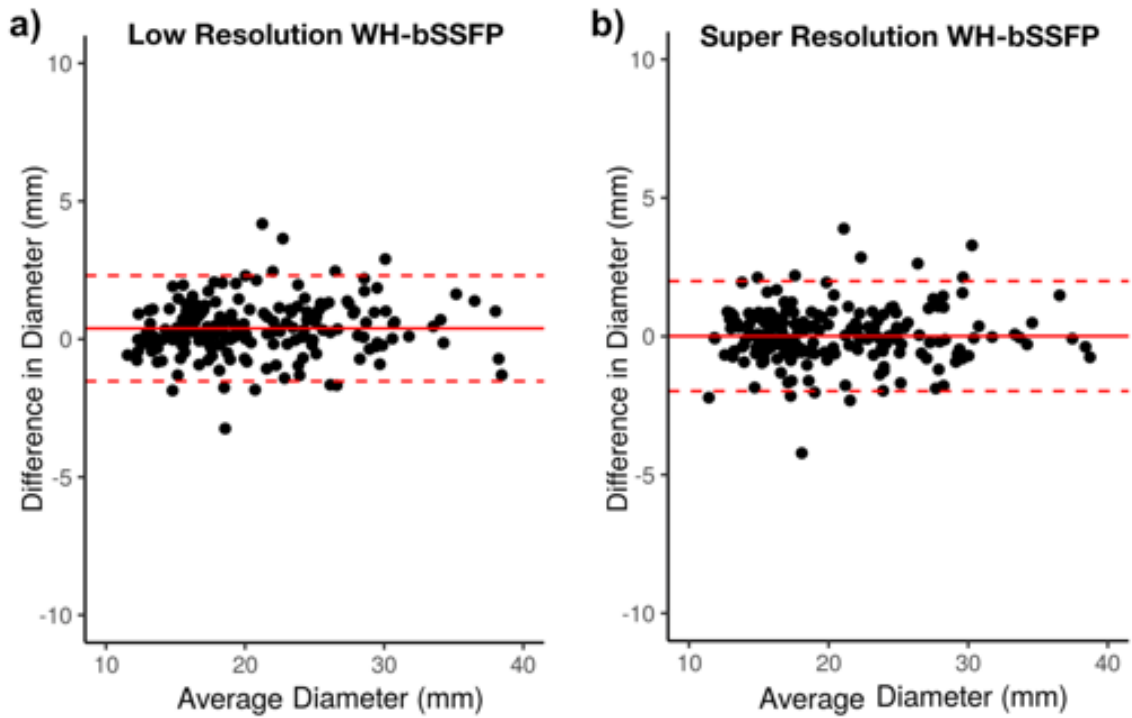

*Figure 6: **Bland-Altman agreement of vessel diameters** Primary observer; Bland-Altman plots of agreement with high-resolution WH-bSSFP for all vessels (see Additional File 5 for the Bland-Altman plots of the individual vessels). The solid red line indicates the bias, with the dashed red lines showing the upper and lower limits of agreement (bias±1.96xStandardDeviation) between the techniques.*

The inter-observer and intra-observer ICC's are shown in Table 3. The largely overlapping confidence intervals demonstrated that there were no significant differences in inter-observer and intra-observer variability between any of the techniques in any of the vessels.

|  | **High-resolution ICC** | **Low-resolution ICC** | **Super-resolution ICC** |
|---|---|---|---|
| *Intra-observer variability* | | | |
| AAo | 0.99 (0.99 to 1.00) | 0.99 (0.97 to 0.99) | 0.99 (0.98 to 1.00) |
| DAo | 0.99 (0.99 to 1.00) | 0.98 (0.95 to 0.99) | 0.99 (0.97 to 0.99) |
| MPA | 0.96 (0.92 to 0.98) | 0.97 (0.93 to 0.99) | 0.97 (0.93 to 0.98) |
| RPA | 0.99 (0.97 to 0.99) | 0.99 (0.98 to 1.00) | 0.99 (0.99 to 1.00) |
| LPA | 0.98 (0.97 to 0.99) | 0.98 (0.96 to 0.99) | 0.98 (0.93 to 0.99) |
| *Inter-observer variability* | | | |
| AAo | 0.97 (0.93 to 0.99) | 0.96 (0.90 to 0.98) | 0.97 (0.94 to 0.99) |
| DAo | 0.84 (0.64 to 0.93) | 0.71 (0.41 to 0.87) | 0.85 (0.66 to 0.94) |
| MPA | 0.93 (0.84 to 0.97) | 0.94 (0.86 to 0.98) | 0.92 (0.82 to 0.97) |
| RPA | 0.78 (0.53 to 0.91) | 0.53 (0.13 to 0.78) | 0.75 (0.47 to 0.89) |
| LPA | 0.87 (0.71 to 0.95) | 0.78 (0.52 to 0.90) | 0.85 (0.66 to 0.94) |

*Table 3: Intra-observer and inter-observer variability* Intra-observer and inter-observer variability; Intra-class correlations for vessel diameters measured from high-resolution, low-resolution and super-resolution WH-bSSFP data. Displayed as ICC (95% confidence intervals).

*Image Quality*

Quantitative and qualitative image quality results can be seen in Table 4. Qualitatively, the low-resolution data was found to have significantly lower sharpness of vessel boarders and more image distortion than the high-resolution data, with no significant difference in residual artefacts. After super-resolution reconstruction, there were no significant differences in terms of qualitative image quality with the high-resolution data. However, a significant improvement was seen in terms of sharpness of vessel boarders and image distortion compared to the low-resolution data.

|  | n | High-resolution | Low-resolution | Super-resolution |
|---|---|---|---|---|
| *Qualitative Image Quality Scores* | | | | |
| Sharpness of vessel borders | 600 | 4.1 ± 0.6 | 3.1 ± 0.7* | 4.2 ± 0.6† |
| Image distortion | 600 | 4.1 ± 0.5 | 3.8 ± 0.5* | 4.0 ± 0.5† |
| Residual artifacts | 600 | 3.8 ± 0.6 | 3.7 ± 0.6 | 3.8 ± 0.5 |
| *Quantitative Image Quality Scores* | | | | |
| SNR | 40 | 17.2 ± 6.5 | 22.8 ± 7.3** | 27.3 ± 11.1**,†† |
| CNR | 40 | 3.0 ± 0.5 | 3.2 ± 0.4** | 3.3 ± 0.4** |
| Edge sharpness (mm$^{-1}$) | 200 | 0.8 ± 0.4 | 0.6 ± 0.3** | 1.3 ± 0.7**,†† |

\* *Indicates significant differences with high-resolution WH-bSSFP technique (p<0.05) as assessed by Friedman's test with post-hoc testing using the Nemenyi test (Qualitative scoring)*

† *Indicates significant differences with low-resolution WH-bSSFP technique (p<0.05) as assessed by Friedman's test with post-hoc testing using the Nemenyi test (Qualitative scoring)*

\*\* *Indicates significant differences with high-resolution WH-bSSFP technique (p<0.05)* ANOVA with post hoc testing using Holm correction *(Quantitative scoring)*

†† *Indicates significant differences with low-resolution WH-bSSFP technique (p<0.05)* ANOVA with post hoc testing using Holm correction *(Qualitative scoring)*

**Table 4** *Qualitative image scores and quantitative image quality results, from the prospective patient study. Displayed as mean ±* standard deviation.

Quantitative analysis showed that the edge sharpness of the low-resolution data was significantly worse than the high-resolution. After super-resolution, the edge sharpness was significantly better than either the low-resolution or high-resolution data. The SNR of the low-resolution data was significantly higher than the high-resolution data. After super-resolution, the SNR improved again, to become significantly higher than either the low-resolution or high-resolution data. The CNR of the three techniques was similar, however the high-resolution technique was found to have be significantly lower than the low-resolution or super-resolution images.

**DISCUSSION**

The main findings of this study were: i) It is possible to train a 3D residual U-Net to perform single volume super-resolution reconstruction on synthetically down-sampled WH-bSSFP data, ii) The accuracy of the network is dependent on the input resolution matching that of the training data, iii) Super-resolution reconstruction of clinically acquired actual low-resolution WH-bSSFP data was successful using the residual U-Net trained using synthetic data, iv) Super resolution data had better image quality than acquired low resolution data and was comparable to reference standard high-resolution data, v) Vessel diameter measurements made using super-resolved data were not significantly different from reference high-resolution data.

**Super-resolution Reconstruction**

The main benefit of super-resolution MR reconstruction is that it can be applied as a post-processing step and therefore, requires no significant sequence modifications. However, conventional super-resolution reconstructions are often computationally intensive and fail to properly recover high resolution features (21, 22). Recently, deep learning has been used to overcome these problems for a range of imaging problems including brain and body MRI (23, 24). In this study, we have developed a deep learning framework for super-resolution of 3D WH-bSSFP data. This was done to speed up acquisition of this time-consuming element of many congenital heart disease CMR protocols.

The main requirement for deep learning is paired input and output data that can be used to train the network. Often this must be prospectively acquired, restricting the ability to quickly develop deep-learning platforms. However, simulating low-resolution data is relatively trivial. Thus, synthetic training data can be easily created from previously acquired high-resolution data, allowing rapid development of this framework. A further advantage of using synthetic data is that the ground truth is known, which allows quantitative evaluation of reconstruction accuracy through measurement of SSIM and MSE. Using these metrics, we were able to show that our network successfully recovers high resolution features from previously unseen synthetic low-resolution data. We also showed that the accuracy of our super-resolution reconstruction was highly dependent on the resolution of the input data.

**In-vivo Study**

Demonstrating reconstruction accuracy on synthetic low-resolution test data is an important first step in framework development. However, for true translation it is vital to test performance on actual clinically acquired low-resolution data. In this study, we successfully used our trained residual U-Net to super-resolve prospectively acquired actual low-resolution WH-bSSFP images. We were able to show that super-resolution reconstruction improved subjective image quality compared to the original low-resolution data. Furthermore, as one might expect, quantitative measures of edge sharpness were higher after super-resolution reconstruction compared to the original low-resolution data. Interestingly, estimated SNR also increased after super-resolution reconstruction, suggesting that the network had some additional de-noising affects.

An important aspect of this study was the comparison of vessels measurements made from high-, low- and super-resolution WH-bSSFP data. We found that vessel diameters were overestimated using the low-resolution data, presumably as a result of the blurred vessel borders. However, there was no statistical differences in vessel diameter measurements between the super-resolution and reference high-resolution data. This suggests that super-resolution reconstruction enabled more accurate vessel measurements to be made from data acquired at low resolution. Importantly, the inter-observer and intra-observer variability of super-resolution reconstruction diameter measurements were similar to high-resolution diameter measurements. This is an important finding as it demonstrates reliability, which is vital for clinical translation.

**Clinical Implications**

We have shown that it is possible to use deep learning super-resolution reconstruction to recover the high-resolution features from low-resolution data. The benefit of acquiring low -resolution data is reduced scan time. In our study, the speed-up in acquisition time between the high-resolution and low-resolution WH-bSSFP was found to be ~x3.0. It should be noted that the resolution was lowered by x2 in both the slice and phase encoding directions, and one might expect a 4x speed up. However, in our implementation the number of GRAPPA reference lines was the same in both the high- and low-resolution acquisitions, slightly limiting the achievable acceleration. Nevertheless, the ability to acquire WH-bSSFP data in less than three minutes is still clinically useful. Importantly, this framework does not require complex sequence modifications, as is necessary for non-Cartesian or compressed sensing optimised acquisitions. This means in theory it is vendor non-specific, as super-resolution reconstruction can be employed as a simple post-processing step. In addition, processing is extremely fast (less than a second per volume) unlike more computationally intensive acceleration techniques, such as compressed sensing. However, we have shown that it is vital that the low-resolution input data matches the synthetically down-sampled data used for training. This currently limits the framework as the way down-sampling is implemented varies between vendor. One solution would be to simply train different networks for different vendor data. Thus, this technique holds the potential to significantly shorten cardiac MR scan times in children.

**Study Limitations**

The main limitation was that we did not assess if super-resolution reconstruction improved the ability to make identify lesions from 3D WH-SSFP data. This is difficult as the heterogeneity of congenital lesions makes it necessary to have a large study population to be adequately powered. However, as this is technique does not require any sequence modification it could easily be disseminated to perform a future multi-centre study.

A further limitation of our approach was that the training and actual input data consisted of coil combined magnitude images, rather than raw multi-coil complex data. The main benefit of this approach was that previously acquired data that was easily retrievable from a conventional clinical image archive could be used for training. However, the absence of phase data in our approach may prevent optimum image restoration.

**Conclusion**

This paper demonstrates the potential of using a residual U-Net for super-resolution reconstruction of rapidly acquired low-resolution whole heart bSSFP data within a clinical setting. Once the network has been trained, the reconstruction times are very short, making these techniques particularly appealing within busy clinical workflow. We have shown that vessel diameter measurements from images reconstructed using a residual U-Net are not statistically significantly different from the reference standard, high-resolution WH-bSSFP techniques. Thus, we believe that this technique may help speed up whole heart CMR in clinical practice.

**LIST OF ABBREVIATIONS**

| | |
|---|---|
| 3D | Three dimensional |
| AAo | Ascending aorta |
| ADAM | Adaptive Moment Estimation algorithm |
| ANOVA | Analysis of variance |
| CMR | Cardiovascular magnetic resonance |
| CNN | Convolutional neural network |
| CNR | Contrast-to-noise ratio |
| DAo | Descending aorta |
| ES | Edge sharpness |
| GPU | Graphics processing unit |
| GRAPPA | GeneRalized Autocalibrating Partial Parallel Acquisition |
| HR | High-resolution |
| ICC | Intraclass correlation |
| LPA | Left pulmonary artery |
| LR | Low-resolution |
| MPA | Main pulmonary artery |
| MPR | Multi-planar reformats |
| MSE | Mean square error |
| RPA | Right pulmonary artery |
| SNR | Signal-to-noise ratio |
| SR | Super-resolution |
| SRR | Super-resolution reconstruction |
| SSIM | Structural similarity index |
| VCG | Vector electrocardiographic gating |
| WH-bSSFP | Whole heart, balanced steady state free precession |

# DECLARATIONS

**Ethics Approval and Consent to Participate**

The local committee of the UK National Research Ethics Service approved the use of retrospectively collected training and test data, as well as collection of prospective whole heart data (06/Q0508/124), and written consent was obtained from all subjects/guardians.

**Consent for Publication**

Permissions: All parents of participants and participants gave consent and assent to participate in the study. Permission was also obtained from parents to publish anonymized patient data collected.

**Availability of Data and Material**

The datasets used and analyzed during the current study are available from the corresponding author on reasonable request.

**Competing Interests**

The authors declare that they have no competing interests.

**Funding**

JAS receives Royal Society-EPSRC funding; Dorothy Hodgkin Fellowship (DH130079). AH is partially supported by the Academy of Finland (Project 312123). This work was supported in part by British Heart Foundation grant NH/18/1/33511.

**Authors' Contributions**

JAS and VM performed study design and were major contributors in writing the manuscript. AH and SA developed the machine learning network architecture. RJ collated much of the training/test data. JAS performed training and testing of the networks, developed OsiriX plugins and performed Edge sharpness, SNR and CNR measurements. VM performed MPR's, as well as calculating statistics. MQ and AG analyzed patient data in terms of diameter measurements and qualitative image scoring. All authors read and approved the final manuscript.

**Acknowledgements**

We would like to express our gratitude to our clinical and research CMR radiographers in Great Ormond Street Hospital, London. This work was supported by the National Institute for Health Research Biomedical Research Centre at Great Ormond Street Hospital for Children National Health Service Foundation Trust and University College London.

# ADDITIONAL FILES

## Additional File 1

*Full demographic information and patient diagnoses.*

|  | Training Data | Synthetic Test Data | Prospective Data |
|---|---|---|---|
| Male/Female | 299/201 | 13/12 | 20/20 |
| Age (years) | 26±13 | 27±12 | 27±14 |
|  | (range: 5-80) | (range: 10-51) | (range: 11-64) |
| Heart rate (bpm) | 67±9 | 69±9 | 68±11 |
|  | (range: 41-86) | (range: 52-85) | (range: 45-95) |
| ***Diagnosis*** | | | |
| Coarctation of the aorta | 57 | 3 | 3 |
| Tetralogy of Fallot / Double outlet right ventricle / Pulmonary atresia with ventricular septal defect | 139 | 3 | 5 |
| Pulmonary valve disease | 33 | 3 | 0 |
| Aortopathy | 81 | 4 | 8 |
| Transposition of the great arteries | 59 | 1 | 5 |
| Aortic valve disease | 20 | 2 | 6 |
| Shunts | 27 | 1 | 6 |
| Complex Congenital Heart Disease | 29 | 1 | 2 |
| Pulmonary Hypertension | 29 | 5 | 4 |
| Tricuspid valve | 11 | 1 | 0 |
| **TOTAL** | **500** | **25** | **40** |

**Additional File 2**

*Flow diagram showing the steps taken to convert the high-resolution WH-bSSFP data, to synthetic low-resolution WH-bSSFP data used to train/test the residual U-Net.*

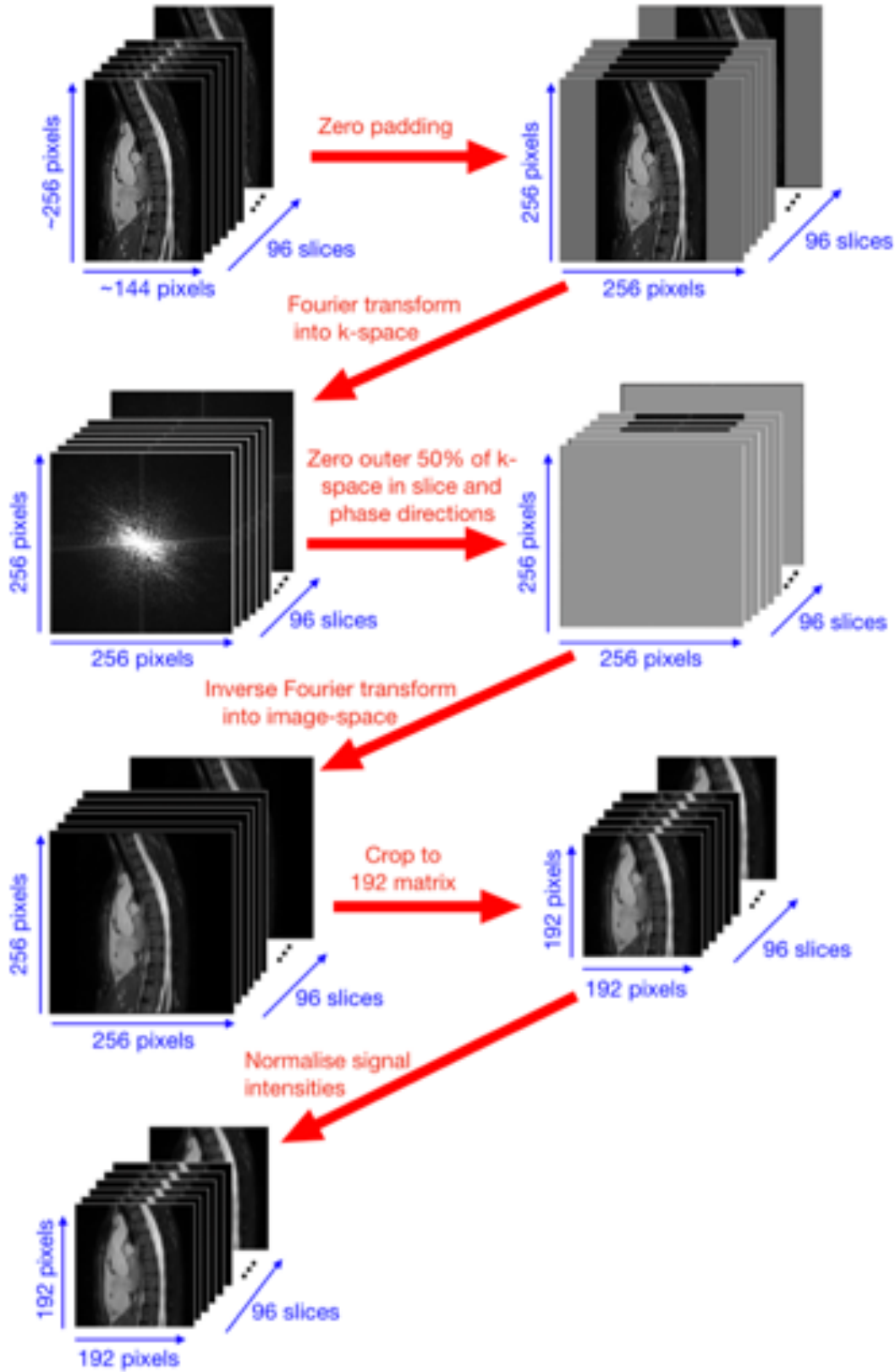

**Additional File 3**

*Synthetic test results from different network structures.*

We investigated two network structures; a U-Net and a residual U-Net, with both a $\ell^1$-loss function and an $\ell^2$-loss function. For each of the four networks, the synthetic training data consisted of 500 paired artefact-free 'ground truth' magnitude images and the corresponding low-resolution images, as described in the main paper.

The resulting networks were tested using 25 previously unseen synthetic low-resolution WH-bSSFP data, as described in the main paper. The table below shows the MSE and SSIM results from the different networks, when compared to the reference high-resolution WH-bSSFP data (mean ± standard deviation over the 25 synthetic test data sets).

|  | SSIM | MSE (x$10^{-3}$) |
|---|---|---|
| Low-resolution data | 0.87±0.02 | 1.28±0.57 |
| Super-resolution data |  |  |
| • U-Net, $\ell^1$-loss | 0.94±0.01* | 0.71±0.45* |
| • U-Net, $\ell^2$-loss | 0.93±0.01* | 0.76±0.46* |
| • Residual U-Net, $\ell^1$-loss | 0.96±0.01* | 0.68±0.45* |
| • Residual U-Net, $\ell^2$-loss | 0.94±0.01* | 0.68±0.44* |

*\*Indicates statistically significantly poorer result compared to the Residual U-Net, $\ell^1$-loss ($p<0.05$)*

*The Residual U-Net with $\ell^1$-loss function had significantly higher SSIM (better reconstruction accuracy) than the other networks ($p<0.05$), with significantly lower MSE (better reconstruction accuracy) than the U-Net with $\ell^1$-loss or $\ell^2$-loss ($p<0.05$). Because of this, the Residual U-Net with an $\ell^1$-loss function was chosen in this paper.*

**Additional File 4**

*Results from the generalisability tests.*

*MSE and SSIM results from the generalisability tests, between high-resolution WH-bSSFP and super-resolved data from different levels of down-sampling. Displayed as mean ± standard deviation over the 25 synthetic test data sets.*

| Percentage of lines Sampled in *ky* and *kz* | SSIM | MSE (x$10^{-3}$) |
|---|---|---|
| 10% | 0.41±0.05* | 16.28±4.44* |
| 20% | 0.62±0.05* | 5.70±1.73* |
| 30% | 0.77±0.04* | 2.52±0.94* |
| 40% | 0.89±0.02* | 1.35±0.67* |
| **50%** | **0.96±0.01** | **0.68±0.45** |
| 60% | 0.95±0.01* | 0.80±0.41 |
| 70% | 0.93±0.01* | 1.14±0.50* |
| 80% | 0.91±0.01* | 1.54±0.55* |
| 90% | 0.91±0.01* | 1.72±0.58* |
| 100% | 0.90±0.01* | 1.91±0.64* |

*Indicates statistically significantly poorer result compared to data sampled with 50% of lines in ky and kz ($p<0.05$)

**Additional File 5**

*Primary observer; Bland-Altman plots of agreement with high-resolution WH-bSSFP for the individual vessels;* ascending aorta (AAo), descending aorta (DAo), main pulmonary artery (MPA), right pulmonary artery (RPA), and left pulmonary artery (LPA). *The solid red line indicates the bias, with the dashed red lines showing the upper and lower limits of agreement (bias±1.96xStandardDeviation) between the techniques.*

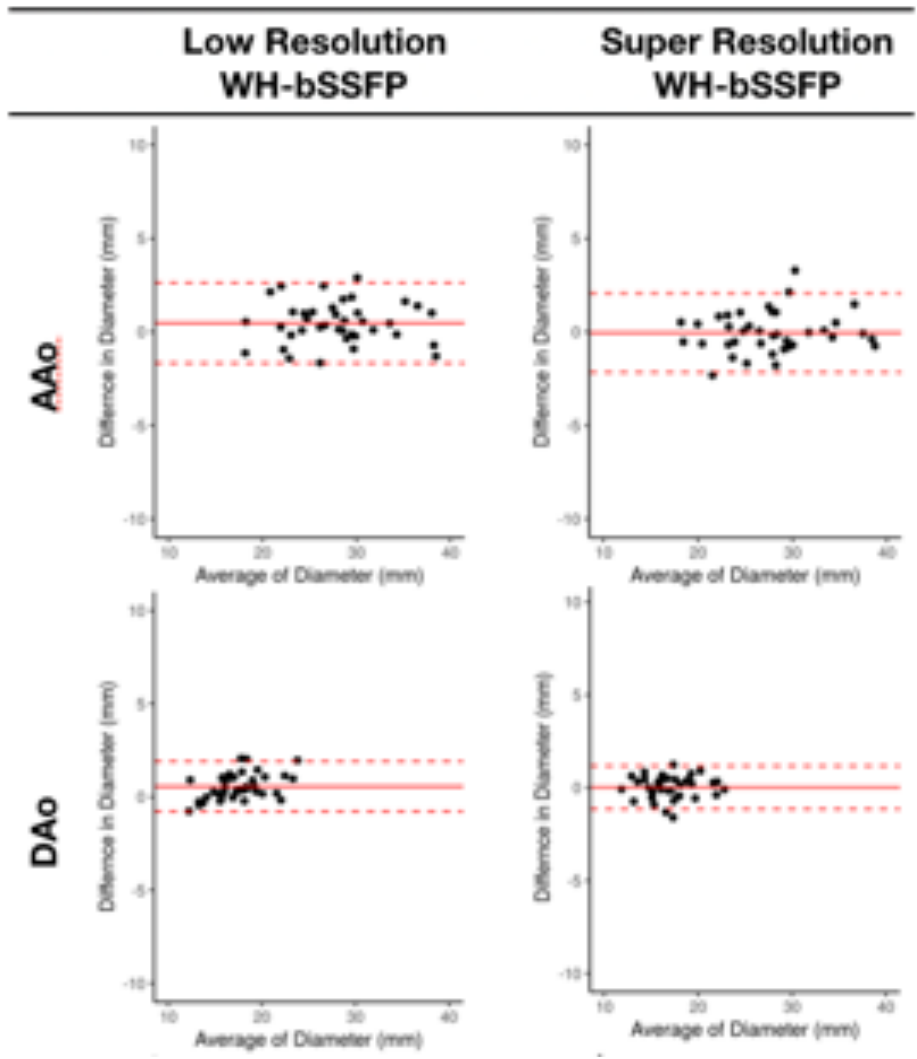

| Low Resolution WH-bSSFP | Super Resolution WH-bSSFP |
|---|---|
| 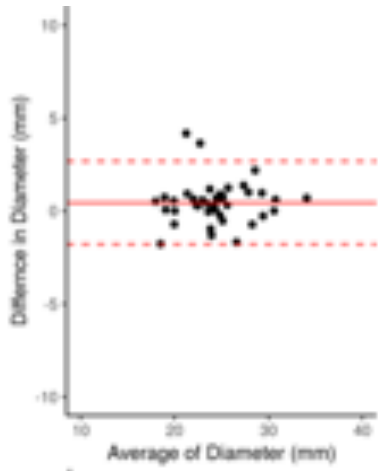 | 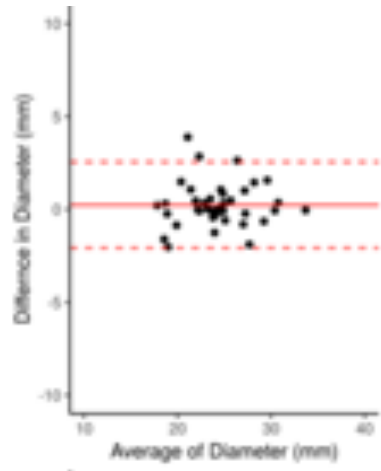 |
| 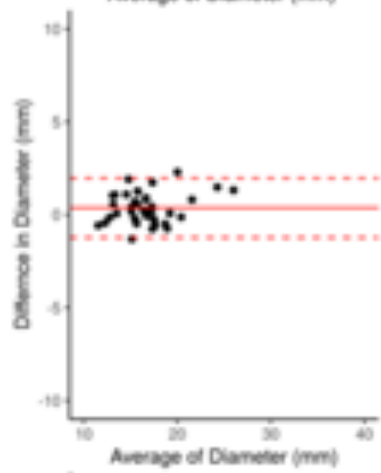 | 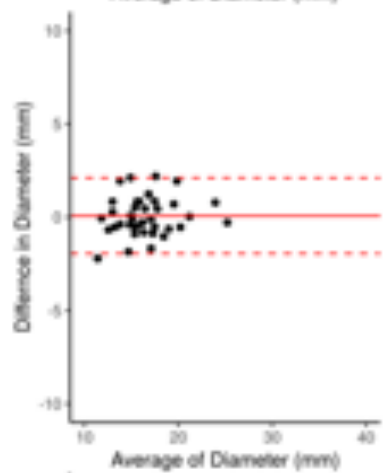 |
| 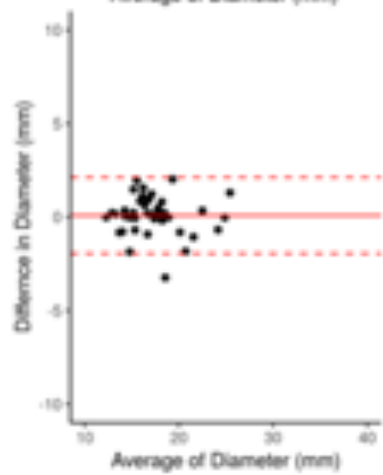 | 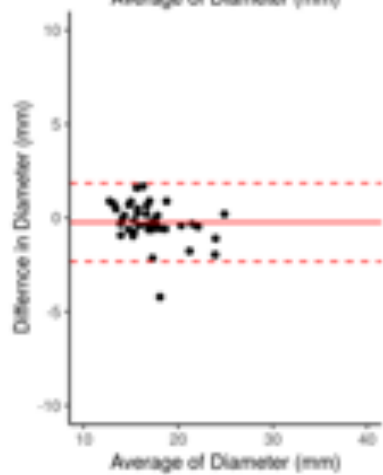 |

MPA (top row), LPA (middle row), RPA (bottom row)